# Spin-fluctuations drive the inverse magnetocaloric effect in Mn$_5$Si$_3$

N. Biniskos,[1,2,*] K. Schmalzl,[1] S. Raymond,[2,†] S. Petit,[3] P. Steffens,[4] J. Persson,[5] and T. Brückel[5,6]

[1]Forschungszentrum Jülich GmbH, Jülich Centre for Neutron Science at ILL, 71 avenue des Martyrs, 38000 Grenoble, France
[2]Univ. Grenoble Alpes, CEA, INAC, MEM, 38000 Grenoble, France
[3]Laboratoire Léon Brillouin, CEA, CNRS, Université Paris-Saclay, CE-Saclay, F-91191 Gif sur Yvette, France
[4]Institut Laue-Langevin, 71 avenue des Martyrs, 38000 Grenoble, France
[5]Forschungszentrum Jülich GmbH, Jülich Centre for Neutron Science (JCNS-2) and Peter Grünberg Institut (PGI-4), JARA-FIT, 52425 Jülich, Germany
[6]Forschungszentrum Jülich GmbH, Jülich Centre for Neutron Science at MLZ, Lichtenbergstr. 1, 85748 Garching, Germany

(Dated: February 21, 2018)

*Inelastic neutron scattering measurements were performed on single crystals of the antiferromagnetic compound Mn$_5$Si$_3$ in order to investigate the relation between the spin dynamics and the magneto-thermodynamic properties. It is shown that among the two stable antiferromagnetic phases of this compound, the high temperature one has an unusual magnetic excitation spectrum where propagative spin-waves and diffuse spin-fluctuations coexist. Moreover, it is evidenced that the inverse magnetocaloric effect of Mn$_5$Si$_3$, the cooling by adiabatic magnetization, is associated with field induced spin-fluctuations.*

Caloric effects are inherent to magnetization processes. The degree of order of a magnetic system determines the number of accessible states, which in turn defines its magnetic entropy. The Maxwell relation $\left(\frac{\partial S}{\partial H}\right)_{T,p} = \left(\frac{\partial M}{\partial T}\right)_{H,p}$ relates the entropy S(H,T) and the magnetization M(H,T) of a system under constant pressure [1]. This shows that any temperature dependence of the magnetization is coupled to an entropy change when varying the magnetic field. This general principle already points to a wealth of possible magneto-thermodynamic effects. Two well-known realizations are the adiabatic demagnetization of paramagnetic (PM) salts to reach sub-Kelvin temperatures [2] and the giant magnetocaloric effect (MCE) observed near room temperature magneto-structural phase transitions [3-5]. This latter effect can be potentially used for magnetic refrigeration applications in daily life as successfully demonstrated in prototype solid state coolers [6].

Alternatively, in some compounds cooling can be achieved by adiabatic magnetization, a less common effect, the inverse MCE. The discovery of large inverse-MCE around room temperature has attracted interest since it provides added flexibility in the material and functional device design [7, 8]. Inverse MCE potentially occurs in any PM to antiferromagnetic (AF) phase transition [9, 10]. However, this effect is usually not large and systems in which the inverse MCE is of interest undergo a first-order magnetic transformation between distinct magnetic phases: AF to ferrimagnetic (FI) transitions [11, 12], AF to ferromagnetic (FM) transitions [13-15] and collinear to non-collinear AF transitions like in Mn$_5$Si$_3$ [16], the subject of this article.

The inverse MCE in Mn$_5$Si$_3$ is observed between a first-order magnetic transformation from a collinear (AF2) to a non-collinear (AF1) AF phase and the magnetic entropy change is ≈3 JK$^{-1}$kg$^{-1}$ for a field change from 0 to 5T [16]. The present article provides new insight into the origin of the inverse MCE in Mn$_5$Si$_3$ by relating it to the spin dynamics observed by inelastic neutron scattering (INS) from single crystals. Our experiments revealed that the low temperature AF1 phase is characterized by well-defined spin-waves, while the higher temperature AF2 phase is characterized by a coexistence of spin-waves and diffuse spin-fluctuations. Furthermore, the application of a magnetic field in the AF1 phase induces spin-fluctuations by restoring the AF2 phase. The entropy release associated with these spin-fluctuations points to their importance in the inverse MCE in $Mn_5Si_3$.

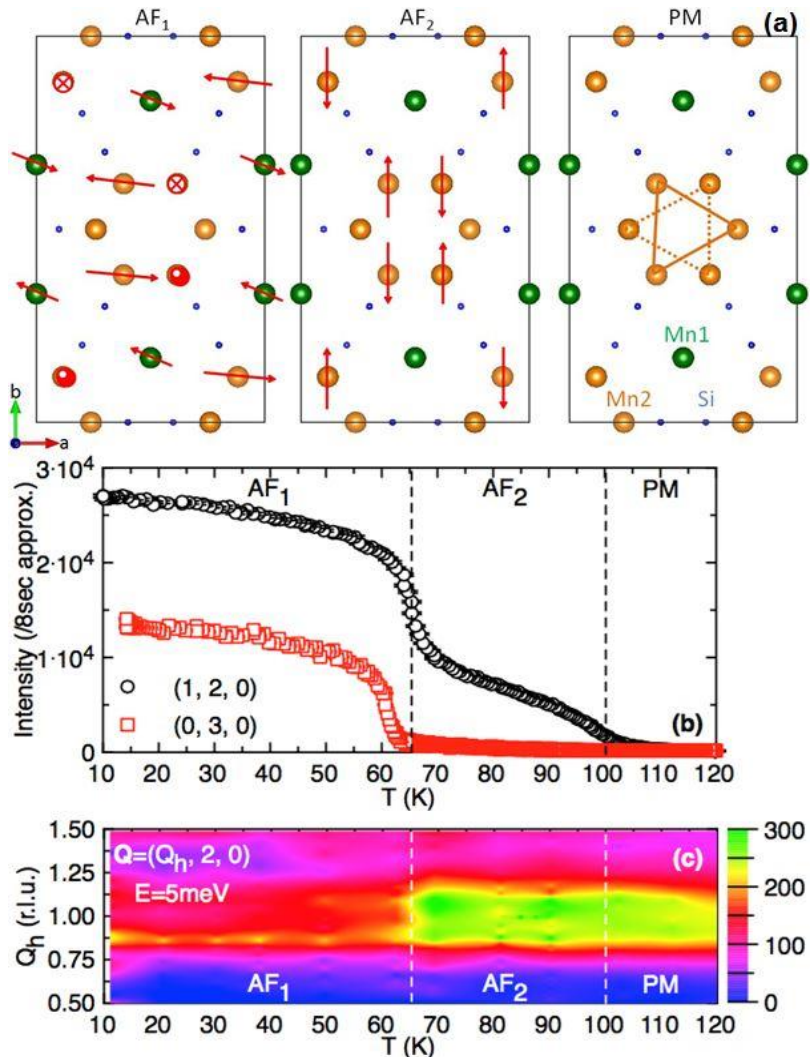

FIG. 1: (a) Projection of the structure of $Mn_5Si_3$ in the ($a,b$) plane of the orthorhombic cell according to single crystal neutron diffraction data [18, 19]. The triangles with continuous and dashed lines are located in different planes separated by $c/2$. The red symbols ⊗ and ⊙ in the AF1 phase indicate magnetic moments pointing in and out of the page, respectively. (b) Temperature dependence of the purely magnetic (1, 2, 0) and (0, 3, 0) Bragg peak intensities. (c) Colour-coded intensity plot of the INS data collected at constant energy transfer of 5meV as a function of $\mathbf{Q}=(Q_h, 2, 0)$ and temperature. Data were collected with unpolarized neutron beam at 2T1. In the inelastic spectra, the background is subtracted and the measured intensity is corrected by the detailed balance factor.

In its high temperature PM state, $Mn_5Si_3$ crystallizes in the hexagonal space group $P6_3/mcm$ with two distinct crystallographic positions for Mn atoms (sites Mn1 and Mn2) [17]. Two first order phase transitions towards antiferromagnetic phases occur at $T_{N2}\approx100$ K (AF2) and $T_{N1}\approx66$ K (AF1), respectively [16]. Associated with the AF2 ordering, a change of crystal structure to the orthorhombic space group $Ccmm$ occurs and Mn2 divides into two sets of inequivalent positions. Magnetic reflections in this cell follow the condition $h+k$ odd corresponding to the magnetic propagation vector **k**=(0, 1, 0) [18]. In this phase, the Mn1 and one third of the Mn2 atoms have no ordered moments and the remaining Mn2 atoms have their magnetic moments aligned almost parallel and antiparallel to the b-axis [18] (see Fig.1(a)). In what follows, AF2 is named as collinear phase. Another structural distortion occurs concomitantly with the AF1 ordering towards an orthorhombic cell without inversion symmetry (space group *Cc2m*) [19]. The magnetic moments reorient in a highly non-collinear and non-coplanar arrangement, while the propagation vector remains the same. Mn1 atoms acquire a magnetic moment and still one third of the Mn2 atoms have no ordered moments, just as in the AF2 phase (see Fig.1(a)). The remaining Mn2 atoms carry a different moment depending on their site. Neutron diffraction [17, 20] and macroscopic measurements [21, 22] indicate that $Mn_5Si_3$ undergoes several magnetic phase transitions under magnetic field and temperature with the inverse MCE observed in the vicinity of the AF1-AF2 phase transition. The non-collinear AF1 phase of $Mn_5Si_3$ is also responsible of the anomalous Hall effect with a switch in electronic transport properties between AF1 and AF2 [21, 23, 24].

INS measurements were performed on single crystal samples; for experimental details see [25]. The temperature dependence of the purely magnetic (1, 2, 0) and (0, 3, 0) Bragg peak intensities are shown in Fig.1(b). (1, 2, 0) marks the onset of AF2 ordering. Neutron scattering is sensitive to magnetic moments or fluctuations perpendicular to the scattering vector. The **Q** vector of (0, 3, 0) is almost parallel to the magnetic moments in the AF2 phase and the associated Bragg peak intensity therefore nearly vanishes. However, this Q position is convenient to detect the AF1 transition. A color-coded intensity map of the INS intensity measured for a constant energy transfer of E=5 meV as a function of **Q**=($Q_h$, 2, 0) and T is shown in Fig.1(c). The PM scattering constitutes of a broad peak centered at the magnetic center $Q_h$ =1 r.l.u., the signature of a correlated diffuse signal. Short range spin correlations persist in the PM state, giving rise to the observed inelastic response. The shape of the scattering of the AF2 phase resembles the one of the PM state, no marked change in the dynamical response occurs at $T_{N2}$. In contrast, AF1 ordering is characterized by a strong modification of the spectrum leading to sharp spin-wave peaks identified through the ridges at $Q_h$ =0.87 r.l.u. and $Q_h$ =1.13 r.l.u., the former being more prominent due to the instrumental resolution focusing conditions. Hence the spin excitation spectrum is markedly different in the two magnetically ordered phases. The same observations stem from the cuts obtained along the ($Q_h$, 2, 0) direction at another energy transfer of E=3meV and shown in Fig.2(d) for selected temperatures in the different phases (cuts along the (0, $Q_k$, 1) and (0, 3, $Q_l$) high symmetry directions are shown in [25]).

To confirm this observation, INS spectra were collected in large portions of the reciprocal space in (*b*,*c*) scattering plane for E=3meV at three different temperatures corresponding to the PM state (T=120 K), the collinear AF2 phase (T=80 K) and the non-collinear AF1 phase (T=10 K). In the AF1 phase, the rings in the Fig. 2(a) represent intense phonon and spin-wave scattering originating from the structural ($h+k$ even) and magnetic ($h+k$ odd) zone centers, respectively. Consistently with Fig.1(c), the spin excitation spectrum is different in the AF1 and AF2 phases for the inelastic spectra that originate from the magnetic Bragg peaks that exist in the two phases (e.g. (0, 3, 1), (0, 1, 1)). While in the AF1 phase clear rings indicate spin-wave scattering (see Fig. 2(a)), the signal in the AF2 phase (bright yellow spots in Fig. 2(b)) resembles the one of the PM state (Fig.2(c)). Additional spectra at different energies are provided in [25] in order to demonstrate the dispersive nature of the spin-waves in the AF1 phase and the diffusive nature of the spin-fluctuations in the PM state.

The spin dynamics of the AF2 phase is very peculiar, it is at first sight not constituted of discrete modes, but of a broad continuum of states. In order to get more insight concerning this behavior and to compare it with the seemingly identical PM spin dynamics, spectra were collected using polarized INS methods. In such an experiment the polarization of the incident neutron beam is consecutively turned into different directions, which gives access to different neutron cross sections. In the present work, the momentum transfer $\hbar\mathbf{Q}$ is defining the x-axis. The initial polarization was prepared parallel to x-axis, perpendicular to $\mathbf{Q}$ in the scattering plane (y-axis) and perpendicular to the scattering plane (z-axis) and the final polarization was analyzed for a scattering process reversing the initial polarization by $180^0$. The corresponding measurement channels are canonically labeled $SF_{xx}$, $SF_{yy}$ and $SF_{zz}$, where SF stands for "Spin-Flip".

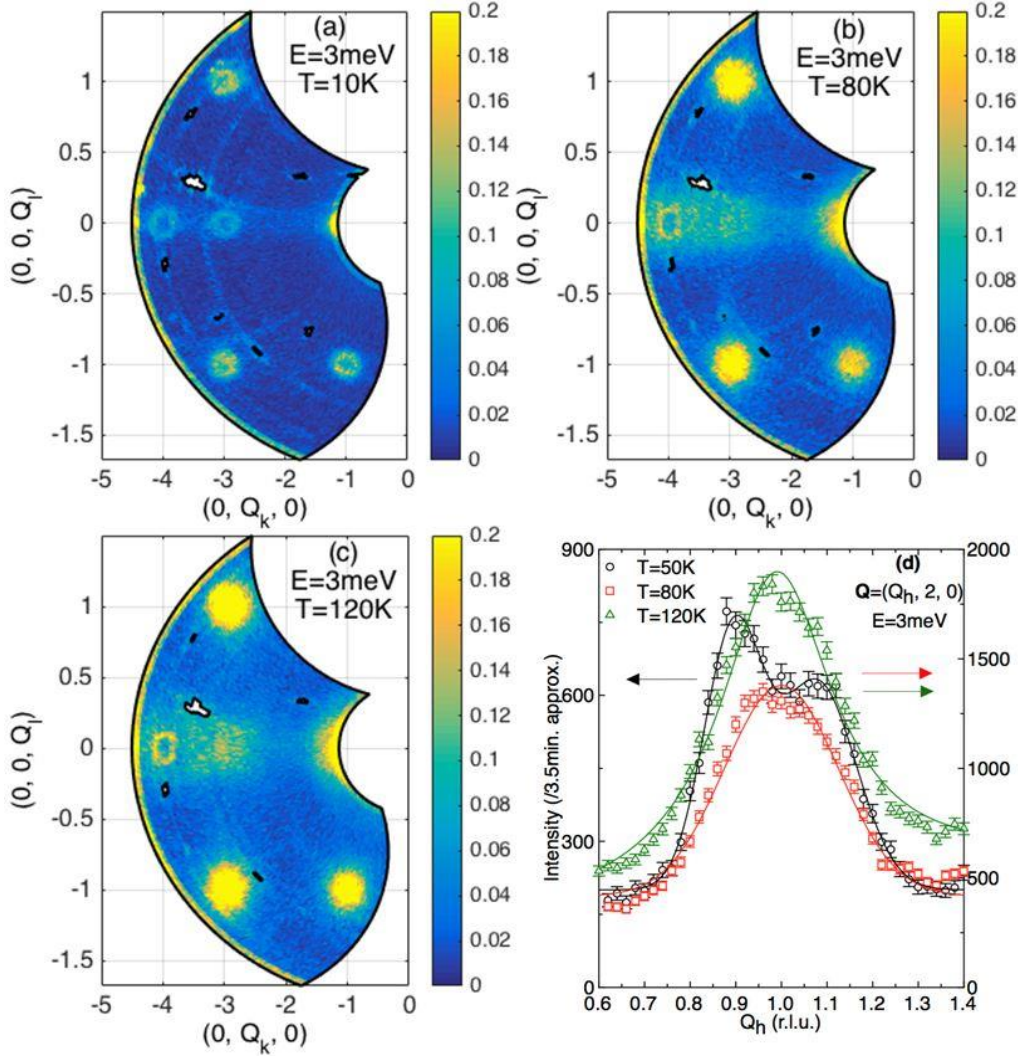

FIG. 2: (a)-(c) FlatCone measurements in the scattering plane (*b,c*) at ThALES in the AF1 (10 K), AF2 (80 K) phase and in the PM (120 K) state at constant energy transfer of 3 meV. The "holes" in the spectra correspond to spurious scattering that has been masked during the data evaluation. (d) Raw data obtained at IN12 with unpolarized neutron beam around $\mathbf{Q}=(Q_h, 2, 0)$ at constant energy transfer of 3 meV. Lines in the AF1 phase (50 K) and in the PM state (120 K) indicate fits with Gaussian and Lorentzian functions, respectively. The line in the AF2 phase (80 K) is a guide for the eyes.

Figures 3 illustrate spectra collected around $\mathbf{Q}=(Q_h, 2, 0)$ at 80K (AF2) and 120K (PM) for $SF_{yy}$ and $SF_{zz}$ channels ($SF_{xx}$ is shown for the AF2 phase in [25]). The spectra in the PM state at 120K are identical for the two polarization channels $SF_{yy}$ and $SF_{zz}$, which indicates isotropic spin-fluctuations. In contrast, the spectra in the AF2 phase at 80K are different for the two polarization channels concerning intensities and lineshapes. This indicates that the magnetic excitation spectrum differs between the AF2 phase and the PM state, a fact that was not evidenced by the unpolarized INS data shown above (see Fig.2(d) and [25]). Moreover, the better wave-vector resolution achieved with the setup used to collect the polarized INS data allows to reveal the flat top shape of the peaks at 80 K. This hints to a signal composed of several ill-resolved peaks. Altogether these information point to the fact that the signal in the AF2 phase is composed of different components.

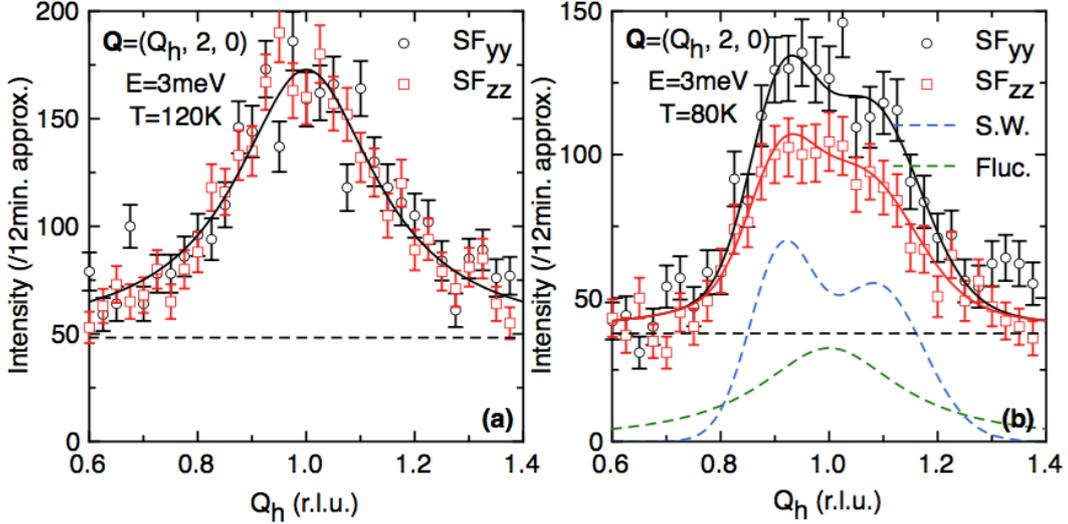

FIG. 3: Inelastic spectra at constant energy transfer of 3meV around $\mathbf{Q}=(Q_h, 2, 0)$ obtained with polarized neutrons in the PM state (120 K) and in the AF2 phase (80 K) at IN12. (a) In the PM state (120 K) the solid line corresponds to a Lorentzian function fit. (b) In the AF2 phase (80 K) the solid lines represent overall fits and the dashed lines the individual signal from spin-waves (S.W.) and spin-fluctuations (Fluc.) (see details in [25]). The horizontal dashed line indicates the background level.

The simplest hypothesis is to consider that it is composed of spin-fluctuations coexisting with spin-waves. This natural assumption stems from the nature of the AF2 magnetic phase with mixed magnetic and non-magnetic sites. It is expected that it sustains both kind of excitations: spin-waves, collective precession of spins around the ordered moment and spin-fluctuations, which correspond to weakly correlated spin relaxation of non-ordered moments. Each of these excitations is the landmark of magnetically ordered and disordered phases, respectively. In order to examine this assumption on quantitative grounds, a simultaneous fit of the data obtained in the two channels $SF_{yy}$ and $SF_{zz}$ was performed. With such a fit the dimensionality of the parameter space is decreased (see equations in [25]). It is furthermore assumed that the spin-fluctuations are isotropic at 80K (as demonstrated at 120 K, See Fig.3a) and that the spin-wave precession is isotropic around the *b*-axis. The resulting fit shown in Fig.3b for 80 K, allows to separate the proportion of spin-waves (blue dashed line) and spin-fluctuations (green dashed line) composing the overall signal. The data are well described by our model and since the polarized INS cross-sections conveys stringent fitting conditions, it gives credit to the hypothesis of the coexistence of spin-waves and spin-fluctuations.

Previous neutron diffraction measurements performed under magnetic field in powder and single crystal samples indicate a transition from the AF1 to the AF2 phase at 3.5T for 58K [17, 20]. The full magnetic phase diagram as a function of temperature and magnetic field up to 10T applied along the c-axis was established by electrical transport and magnetization measurements [21] and is sketched in Fig.4a. Below 60 K, the increasing magnetic field induces transitions from the AF1 to another intermediate AF1' phase before reaching the AF2 phase. Above 60 K, the AF2 phase is stable up to the maximum investigated field of 10 T.

Building up on these results, we investigated the field dependence of the magnetic Bragg peak intensities (1, 2, 0) and (0, 3, 0) at T=50K for a field applied along the *c*-axis (See Figs 4 (b) and (c)). Starting at H=0T in the AF1 phase, two anomalies at approximately 3 and 7 T mark the onset of different magnetic phases that are identified as AF1' and AF2 phases, roughly consistent with the (T-H) phase diagram of Ref. [21]. The intermediate AF1' phase is not addressed in the present study. Above 7 T, the system re-enters into the AF2 phase characterized by a drop of a factor ≈2 compared to AF1 in the intensity of the Bragg peak (1, 2, 0) and a continuous diminution of the (0, 3, 0) Bragg peak intensity. Up to the maximum investigated field of 10 T, the PM state is not reached, which is related to the very steep $T_{N2}(H)$ phase boundary. The field induced AF2 phase at 50K consists of antiferromagnetically coupled magnetic moments lying perpendicular to H with an additional increasing induced ferromagnetic component, which is evidenced in magnetization measurements [21].

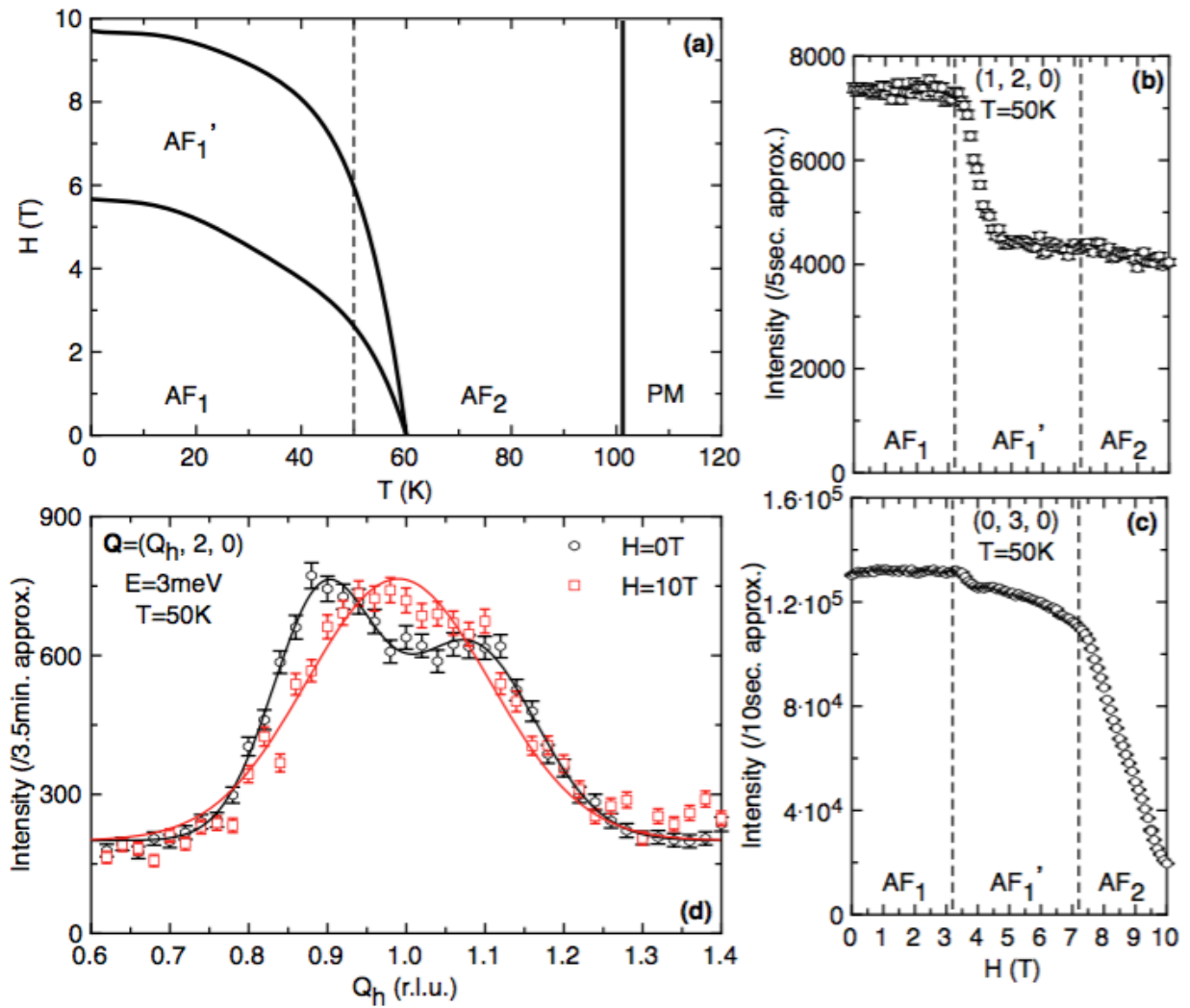

FIG. 4: (a) (T-H) phase diagram of $Mn_5Si_3$ for **H**∥*c* after Ref. [21]. The vertical dashed line indicates the temperature (50 K) where neutron scattering measurements were performed under magnetic field. (b)-(c) Field dependence of magnetic Bragg peaks (1, 2, 0) and (0, 3, 0) at 50 K. For (1, 2, 0), a 9mm plexiglass attenuator was in place. (d) Inelastic spectra obtained at 50K around **Q**=($Q_h$, 2, 0) under 0T and 10T magnetic field for an energy transfer of 3 meV. The red line is guide for the eyes and the black line correspond to fits with Gaussian functions. Data were obtained with unpolarized neutron beam at IN12.

Fig.4(d) shows INS spectra collected around **Q**=($Q_h$, 2, 0) at 50K at a constant energy transfer of E=3meV for H=0 and 10 T. The typical spin-wave spectra with two peaks in the AF1 phase at H=0T transforms into a broad peak centered at the magnetic center $Q_h$ =1 r.l.u. at 10 T. Drawing a parallel between the spectra at 50K as a function of field (Fig.4d) with the ones at H=0T as a function of temperature (Fig.2d), one concludes that the field restores the fluctuations pattern associated with the AF2 phase. This leads to the unusual feature that a high enough magnetic field of 10T applied in the AF1 phase induces spin-fluctuations. In contrast, a field of 10T applied in the AF2 phase and PM state reduces the intensity of the magnetic excitations without changing their lineshapes (see [25]).

The microscopic ingredients giving rise to the stability of the mixed magnetism of the AF1 and AF2 phases, constituting of the coexistence of magnetic and non-magnetic sites, were discussed in previous studies on $Mn_5Si_3$ [17-21]. Namely, a first ingredient at play is the Mn1-Mn1 distance with respect to the critical value corresponding to the instability in the competition between bonding and magnetism. The shortest Mn1-Mn1 distance in $Mn_5Si_3$ equals $c/2$ and is too small in the AF2 state to lead to a magnetic configuration for Mn1. Below $T_{N1}$, an abrupt change of the lattice parameter value c occurs and this allows to stabilize the magnetic configuration of the Mn1 site. Under magnetic field, it was shown by neutron diffraction experiment performed at 50K and 4T [17], that $c/2$ decreases below the critical value for Mn1 moment stability. A second ingredient common to the AF1 and AF2 phases is the geometrical frustration associated with the triangular arrangement of Mn2 atoms, which leads to the occurrence of two ordered moments out of the three Mn2 sites (See Fig.1(a)).

Such mixed magnetic phases occur in other Mn-based compounds. The most extensively studied ones are the Laves phases $RMn_2$ (R being a rare earth element) for fundamental magnetic properties [26], as well as, for the MCE [27]. In $RMn_2$ systems, the same ingredients as in $Mn_5Si_3$ are at play: instability of Mn magnetism and magnetic frustration. The switch from conventional to mixed AF phase as a function of magnetic field was theoretically predicted and verified by neutron diffraction experiments in $TbMn_2$ [28]. This situation bears similarity to $Mn_5Si_3$ and our results find echo in the broader context of mixed magnetic systems studies. The present work goes one step further by evidencing the properties of the excitation spectrum of the mixed magnetic phases revealing the coexistence of two dynamical responses, spin-waves and spin-fluctuations.

Another important result of our study is the evidence of magnetic field induced spin-fluctuations. Starting from the AF1 phase, which sustains only spin-waves, the field restores the AF2 phase and its associated spin-fluctuations. In contrast to the discrete spin-wave spectrum, such spin-fluctuations which consist of a continuum of excitations involve more microscopic states. This leads to an increase of the magnetic entropy and thus plays a major role in the inverse MCE. It is to be noted that the observed behaviour is opposite to the most common magnetic field effect that usually suppresses the spin-fluctuations and consequently decreases the magnetic entropy. The latter leads to the direct MCE as recently demonstrated by INS in the FM MCE compound $MnFe_4Si_3$ [29].

In summary, we evidence the coexistence of spin-waves and spin-fluctuations in the high temperature mixed AF phase of $Mn_5Si_3$. We show that the modification of the magnetic excitation spectrum induced by the magnetic field tracks the inverse MCE via the associated change of magnetic entropy. The need of designing functional materials for caloric applications connects with a fundamental understanding of magnetism.

*Electronic address: n.biniskos@fz-juelich.de
[†]Electronic address: raymond@ill.fr

## SUPPLEMENTAL MATERIAL

### S1. Synthesis

For synthesizing polycrystalline samples of Mn$_5$Si$_3$, elementary manganese (Aldrich, 99.99%) and silicon (Aldrich, 99.99 %) were used. The elements were mixed in stoichiometric ratios and melted in argon atmosphere by induction heating in a levitation cold crucible. The resulting product was cooled and heated four times to ensure maximum homogeneity of the sample. In order to confirm the formation of the Mn$_5$Si$_3$ phase, part of the polycrystalline sample was characterized by X-ray powder diffraction at room temperature. Once the formation of the Mn$_5$Si$_3$ phase was confirmed, a single crystal was grown from the polycrystalline samples using the Czochralski method in an aluminum oxide crucible with a tungsten crystal as a seed.

### S2. INS experiments

Two single crystals (with mass of about 7 g each) were individually mounted on an aluminium sample holder and oriented in the (*a*,*b*) and (*b*,*c*) scattering plane of the orthorhombic symmetry, respectively. The linewidths of the rocking curves of the samples consist of single Gaussian peaks of about $0.25^0$ as determined by neutron diffraction. In order to reduce the scattering of the aluminium sample holder a cadmium foil was placed around it.

INS measurements were carried out on the cold triple axis spectrometers (TAS) ThALES and IN12 at the Institut Laue Langevin (ILL) in Grenoble, France, as well as on the thermal TAS 2T1 at the Laboratoire Léon Brillouin (LLB) in Saclay, France. All spectrometers used for INS studies were setup in W configuration with a fixed final energy and focusing setups were employed. Additional information regarding each configuration is given in Table I.

For unpolarized INS measurements below room temperature, the sample was cooled down with a $^4$He flow cryostat (for TAS at ILL) or a closed cycle cryostat (for TAS at LLB), covering the temperature region of 10≤T≤120 K. In order to map extended ranges of the (**Q**, E)-space, configuration "A" with the FlatCone multianalyzer option was used [1]. The rotation step width of the sample was $0.5^0$ and the measuring time per step 2 min.

Spin dynamics investigations with unpolarized neutrons under magnetic field were carried out using a 10T vertical field magnet. The single crystal was oriented in the (*a*,*b*) scattering plane of the orthorhombic structure. The field was applied along the $c$-axis (configuration "B" in Table I) and 80'-open-open collimations were installed.

Polarized INS measurements were performed with configuration "D" in Table I, using the spherical polarization analysis setup CRYOPAD [2]. The incident neutron beam spin state was prepared with a transmission polarizing cavity located after the velocity selector [3]. All along the neutron beam path guide fields were installed to maintain the polarization of the beam. A flipping ratio of about 14 has been determined from a graphite sample.

In this article, we use the ortho-hexagonal representation of the cell with lattice parameters $a$=6.88 Å, $b$=11.91 Å and $c$=4.81 Å at 10 K. The scattering vector **Q** is expressed in Cartesian coordinates **Q**=($Q_h$, $Q_k$, $Q_l$) given in reciprocal lattice units (r.l.u.). The wave-vector **q** is related to the momentum transfer through $\hbar\mathbf{Q}=\hbar\mathbf{G}+\hbar\mathbf{q}$, where **G** is a Brillouin zone center and **G**=($h$, $k$, $l$).

TABLE I: Instrument configurations.
"PG" refers to pyrolytic graphite. Higher order contamination was removed using a velocity selector (VS) before the monochromator on ThALES and IN12 and a PG filter in the scattered neutron beam on 2T1. The symbol "*" refers to polarized setup.

| Config. | TAS | Monoch. | Anal. | $k_f$ (Å$^{-1}$) | Filter |
|---|---|---|---|---|---|
| A | ThALES | Si(111) | Si(111) | 1.4 | VS |
| B | IN12 | PG(002) | PG(002) | 1.8 | VS |
| C | 2T1 | PG(002) | PG(002) | 2.662 | PG |
| D | IN12* | PG(002) | Cu$_2$MnAl(111) | 1.8 | VS |

## S3. INS spectra along the $b$ and $c$ directions in the two AF phases and in the PM state

Cuts obtained at the energy transfer of E=3meV for selected temperatures corresponding to the different phases along the (0, $Q_k$, 1) (a) and (0, 3, $Q_l$) (b) directions are shown in Fig.S1. The shape of the scattering of the AF2 (80 K) phase resembles the one of the PM state (120 K). Clear peaks in the AF1 phase (10 K) at finite $q$ away from the magnetic center correspond to spin-wave scattering. These results are consistent with the FlatCone data shown in the main text. It is demonstrated that the spin excitation spectrum is different in the two magnetically ordered phases along all the three orthorhombic high symmetry directions, namely ($Q_h$, 2, 0) (see Fig.2(d) in the main text), (0, $Q_k$, 1) Fig.S1(a) and (0, 3, $Q_l$) Fig.S1(b).

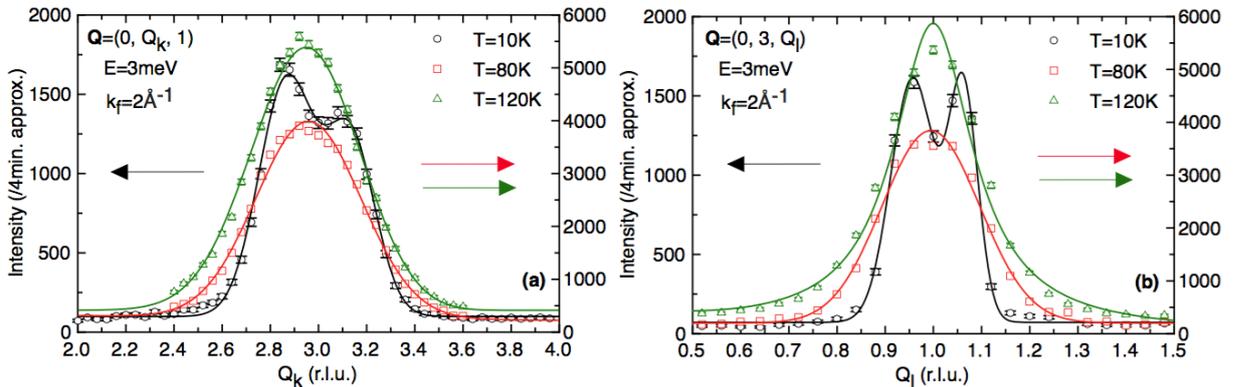

FIG. S1: Spectra obtained at IN12 with unpolarized neutrons around **Q**=(0, $Q_k$, 1) (a) and **Q**=(0, 3, $Q_l$) (b) in the AF1 (10 K), AF2 (80 K) phase and in the PM (120 K) state at constant energy transfer of 3 meV. Lines in the AF1 phase (10 K) and in the PM state (120 K) indicate fits with Gaussian and Lorentzian functions, respectively. The line in the AF2 phase (80 K) can be used as guide for the eyes.

## S4. Propagating spin-waves of AF1 phase and diffusive spin-fluctuations of PM state

Figs S2(a)-(c) show spectra at different constant energy transfers in the AF1 phase (10 K) along the three high symmetry orthorhombic directions. The spectra show a maximum at finite $q$ away from the magnetic zone center, which shifts to higher values with increasing energy transfers. This behavior indicates that the peaks correspond to dispersive spin-waves. On the contrary, for the spectra in the PM state (120 K) (see Fig.S2(d)) there is no well-defined inelastic mode associated with a given wave vector, all spectra are centered at the magnetic zone center and the intensity is decreasing with increasing energy transfer, while the signal broadens. This is typical for PM scattering and indicates the diffusive nature of the spin-fluctuations.

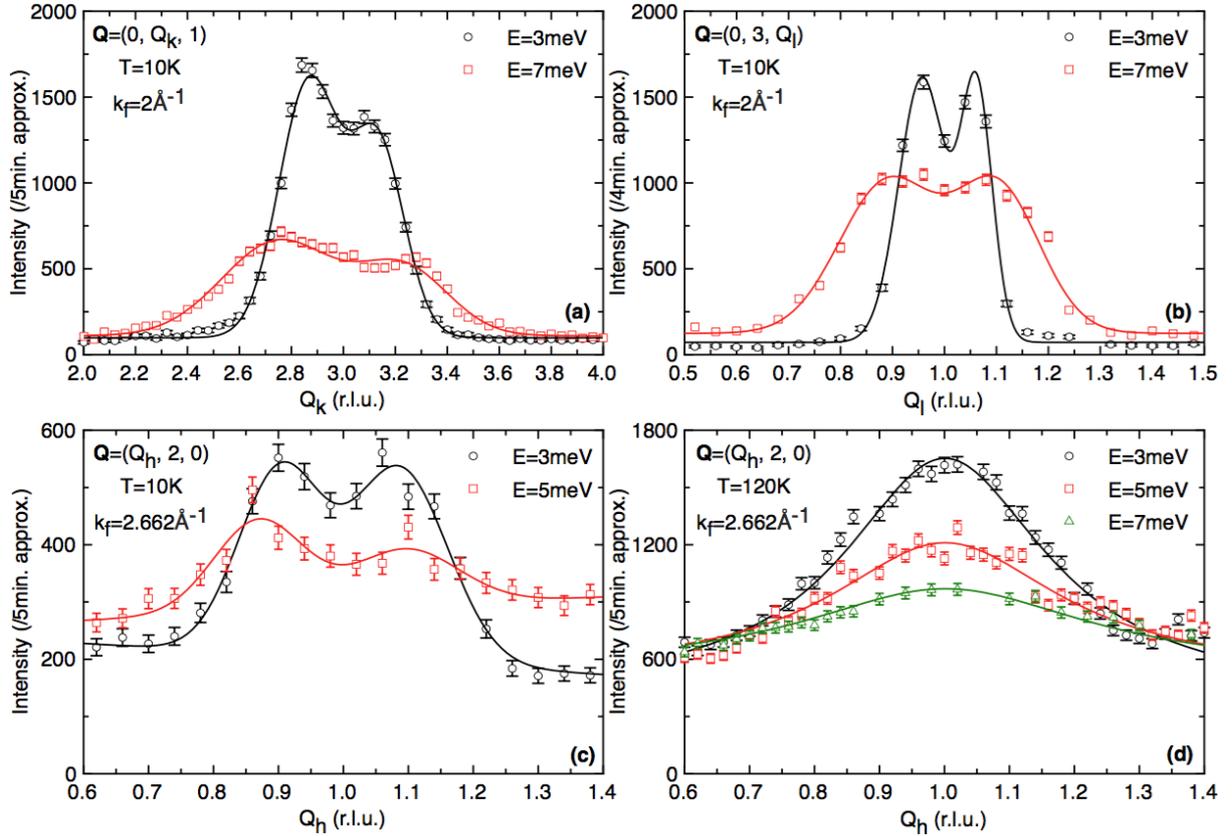

FIG. S2: (a)-(c) Raw data depicting spin-wave scattering at different constant energy transfers along the $(0, Q_k, 1)$ (a), $(0, 3, Q_l)$ (b) and $(Q_h, 2, 0)$ (c) high symmetry direction of the orthorhombic system at 10K (AF1 phase). Lines represent fits with Gaussian functions. (d) PM scattering at different constant energy transfers along the $(Q_h, 2, 0)$ high symmetry direction of the orthorhombic system at 120 K. Lines represent fits with Lorentzian functions. Panels (a)-(b) and (c)-(d) show IN12 and 2T1 data, respectively.

## S5. Polarized INS experiments

For polarized INS, a Cartesian coordinate system is defined with the x axis parallel to the momentum transfer $\hbar \mathbf{Q}$, the y axis perpendicular to $\mathbf{Q}$ in the scattering plane, and the z axis perpendicular to the scattering plane. Neutron scattering experiments probe only the magnetism perpendicular to the scattering vector $\mathbf{Q}$, the measured magnetic fluctuations are $\langle \delta M_y \rangle$ and $\langle \delta M_z \rangle$ [4]. Magnetic fluctuations parallel to the neutron beam polarization are visible in the non-spin-flip channel, while fluctuations perpendicular to the neutron beam polarization give rise to the spin-flip signal (i.e. inversion of the polarization of the scattered beam with respect to the initial one).

In order to have a quantitative result concerning the spin-wave and the spin-fluctuation contributions to the spectra several assumptions were made. Assuming that the background is the same for the three spin-flip channels ($BG_{SF}$), the neutron scattering double differential cross sections are:

$$SF_{xx} : \left(\frac{d^2\sigma}{d\Omega dE}\right)_{SF}^{x} \propto BG_{SF} + \langle \delta M_y \rangle + \langle \delta M_z \rangle \quad (1)$$

$$SF_{yy} : \left(\frac{d^2\sigma}{d\Omega dE}\right)_{SF}^{y} \propto BG_{SF} + \langle \delta M_z \rangle \quad (2)$$

$$SF_{zz} : \left(\frac{d^2\sigma}{d\Omega dE}\right)_{SF}^{z} \propto BG_{SF} + \langle \delta M_y \rangle \quad (3)$$

In the crystal frame and for **Q** in the ($a$,$b$) plane, the cross sections become:

$$SF_{yy} : \left(\frac{d^2\sigma}{d\Omega dE}\right)_{SF}^{y} \propto BG_{SF} + \langle \delta M_c \rangle \quad (4)$$

$$SF_{zz} : \left(\frac{d^2\sigma}{d\Omega dE}\right)_{SF}^{z} \propto BG_{SF} + \cos^2\theta \langle \delta M_b \rangle + \sin^2\theta \langle \delta M_a \rangle \quad (5)$$

with $\theta$ the angle between **Q** and the [100] direction and can be calculated by $\theta = \arctan\left(\frac{2}{Q_h}\frac{a}{b}\right)$ for **Q**=($Q_h$, 2, 0). In the AF2 phase the magnetic moments lie parallel and antiparallel to the $b$-axis. Spin-waves correspond to precession perpendicular to the ordered moment so their components will appear in $\langle \delta M_c \rangle$ and $\langle \delta M_a \rangle$. Assuming that the signal in this phase consists of spin-waves and spin-fluctuations, the cross sections can be rewritten as:

$$SF_{yy} : \left(\frac{d^2\sigma}{d\Omega dE}\right)_{SF}^{y} \propto BG_{SF} + \langle \delta M_c^{sw} \rangle + \langle \delta M_c^{f} \rangle \quad (6)$$

$$SF_{zz} : \left(\frac{d^2\sigma}{d\Omega dE}\right)_{SF}^{z} \propto BG_{SF} + \cos^2\theta \langle \delta M_b^{f} \rangle + \sin^2\theta \langle \delta M_a^{f} \rangle + \sin^2\theta \langle \delta M_a^{sw} \rangle \quad (7)$$

Assuming that the spin-waves and the spin-fluctuations are isotropic, equations (6) and (7) can be simplified further to:

$$SF_{yy} : \left(\frac{d^2\sigma}{d\Omega dE}\right)_{SF}^{y} \propto BG_{SF} + \langle \delta M^{sw} \rangle + \langle \delta M^{f} \rangle \quad (8)$$

$$SF_{zz} : \left(\frac{d^2\sigma}{d\Omega dE}\right)_{SF}^{z} \propto BG_{SF} + \langle \delta M^{f} \rangle + \sin^2\theta \langle \delta M^{sw} \rangle \quad (9)$$

The measured intensity for the spin-fluctuations and the spin-waves can be described by a Lorentzian and a set of two Gaussians, respectively. In order to reduce the number of fitted parameters, we assume that the Lorentzian function is centered at $Q_h$=1 r.l.u. and that its width equals to the one obtained from the fitting in the PM state (see Fig.3(a) in main text). The resulting fits for the AF2 phase are shown in Fig.3(b) in the main text for $SF_{yy}$ and $SF_{zz}$ and in Fig.S3 for $SF_{xx}$. Consistently the spin-wave scattering in the polarization channel $SF_{zz}$ is reduced due to the angle prefactor $\sin^2\theta$, which decreases with increasing **Q**. It is to be noted that $SF_{yy}$ and $SF_{zz}$ are fitted simultaneously in order to reduce the space of accessible parameters.

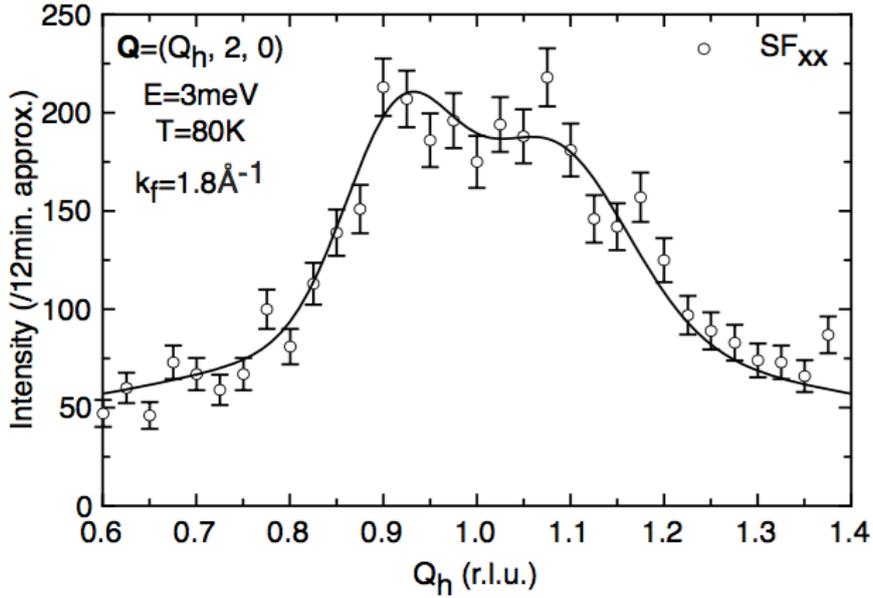

FIG. S3: Inelastic spectrum obtained with polarized neutrons at IN12 in the $SF_{xx}$ channel at 80K around $\mathbf{Q}=(Q_h, 2, 0)$ at constant energy transfer of 3 meV. Line represents a fit as explained in the text.

### S6. Magnetic field measurements in the AF2 phase and in the PM state

Figs S4(a)-(b) show spectra obtained at the energy transfer of E=3meV around $\mathbf{Q}=(Q_h, 2, 0)$ in the AF2 phase (80 K) and in the PM state (120 K). A magnetic field of 10T (Fig.S4(a)) and 8T (Fig.S4(b)) reduces the intensity of the magnetic excitation spectrum in the AF2 phase and PM state, respectively, without changing the lineshapes.

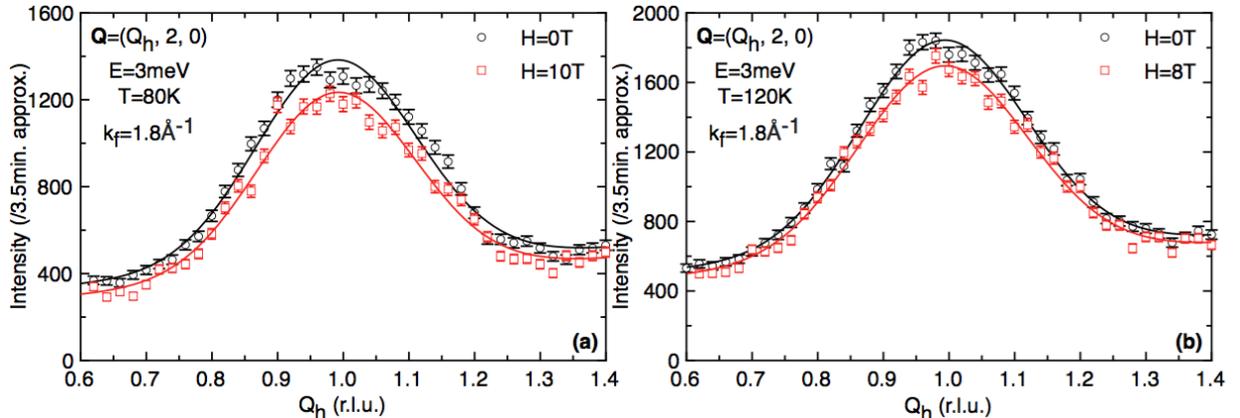

FIG. S4: Inelastic spectra obtained with unpolarized neutrons at IN12 at 80K (a) and 120K (b) around $\mathbf{Q}=(Q_h, 2, 0)$ for E=3meV energy transfer. Data points in black and red indicate measurements with zero and high magnetic field, respectively. The lines are guides for the eyes.